# Self-consistent analysis of the contact phenomena in low-mobility semiconductors

Yevgeni Preezant and Nir Tessler[1]

Electrical Engineering Dept., Technion Israel institute of technology, Haifa 32000, Israel

**Abstract**

Self-consistent solution of charge injection and charge transport in low mobility LEDs is reported. We show that explicit description of the contact region under the same premise as the transport equations is needed to accurately evaluate the current-voltage characteristics of polymer or small-molecule based LEDs. The results are compared to widely used models, which treat the contact region in an implicit manner.

---

[1] E-mail: nir@ee.technion.ac.il
Web page: www.ee.technion.ac.il/nir

# I.  Introduction

Charge injection and transport phenomena have been studied for many years and in many material systems [1] including that of organic semiconductors [5-2]. Many of these studies are now being revisited[10-4,6], as high quality devices seem to emerge through the use of new and better materials[11]. Lately, it has become evident that a better description of the contact region or the contact phenomena in organic-material based devices is required. It has been proposed that one may need to add interface states in the form of traps or dipoles to better simulate experimental results [12,13]. However, it has also been suggested that the contact phenomena in organic-materials should be formulated in a manner adequate to low-mobility semiconductors [14,15] and not as a correction to the contact phenomena in ceramic semiconductors. The common feature of all the models described above is that they treat the contact-phenomena separately or lump the contact-region into a single point in space. Recently, a molecular-oriented transport model[16] that treat the contact region in an explicit manner has been developed and applied to various light emitting diode (LED) structures. In this paper we also make the contact region an explicit part of the device and solve the entire device using a single semiconductor device-model. Moreover, we show that the effect of disorder and the Gaussian density of state (DOS) can be entered into such a model in an easy to implement manner.

# II. The Physical Picture

Before describing the complete model we first examine the physical picture we use to describe the contact region. Investigation and development of physical model for charge injection into organic as well as disordered materials can be traced back by several decades. Microscopic description of charge injection would include ballistic motion of charge carrier through the polymer, energy loss and thermalization, hopping motion of thermalized carrier between the localization site to the collector or recombination at the source. However, a macroscopic semiconductor device-model is generally applied to thermalized carriers only and hence if we want to include the contact region in such a model the thermalization length should be negligibly small. Thermalization of carriers in polymers can be described as ballistic motion of the particle under the influence of viscous drag force in the image potential field [17] :

$$m\frac{dv}{dt} = -e\frac{v}{\mu} - e\frac{d\varphi}{dx} \qquad (1)$$

Which lead to an approximate expression for the thermalization distance for hot carriers:

$$x_t \approx \mu v_0 (m/e) \qquad (2)$$

Where $v_0$ is the initial velocity of the injected carrier, µ is the mobility, and m is the carrier mass. This equation illustrates the relevance of the low-mobility to the physical picture. Although the mass of a carrier-polaron is not well known the overall thermalization length is believed to be in the range of 1- 0.1 nm and that one can assume that the carriers thermalize at first-hop site [15] and any further motion of the carrier is governed by hopping transport in the electronic potential. This process can be modeled by Monte- Carlo simulation [15] or by drift-diffusion equation [17-19].

In the current context one should compare the thermalization length with the size of the contact region, defined as the space between the metal/semiconductor interface and the potential peak (see Figure 1). The length of this contact region varies between ~10nm and ~5nm for applied voltage between 2.5V and 4V, respectively (assuming built-in voltage of 2V and a 100nm thick device). At the low voltage range (relevant to LEDs) the thermalisation length is much smaller then the contact region with the latter comprising a sizeable fraction of the device. Namely, the contact region should be treated in an explicit manner and it can be treated under the premise of drift-diffusion models.

# III. Device Models

## III.A. Space charge limited current

The upper bound of any undoped device is given by so called space charge limited current relation (bulk limited):

$$J_{SCL} = \frac{9}{8}\varepsilon\varepsilon_0\mu\frac{V^2}{L^3} \tag{3}$$

## III.B. Emission Diffusion (generalized SCLC)

While this expression is valid for low barrier injection contacts it can be extended to include contact-limiting effects using the following formulae [1]:

$$J_{ED} = qN_0\mu E(0)\exp(-\frac{q\phi_b}{kT})$$
$$\phi_b = \phi_{b0} - \sqrt{\frac{qE(0)}{4\pi\varepsilon\varepsilon_0}} \tag{4}$$

where $\Phi_b$ is the potential value at its maximum (point $x_m$ in Figure 1). The voltage drop between $x_m$ and the other contact (x=L) is then given by:

$$V_{ED} = \int_0^L \sqrt{E^2(0) + \frac{2Jx}{\mu \varepsilon \varepsilon_0}} dx$$

(5)

The physical picture of this model is transport of charge carriers under the combined (joint) potential of the image force lowered by the applied potential (see Figure 1). If the initial concentration of the carriers at the lower point of potential, at the metallurgic junction, is equal to the total DOS ($N_0$) then the concentration at the top of the potential ($x_m$) is given by:

$$n(x_m) = N_0 \exp(-\frac{q\phi_b}{kT})$$                      (6)

At $x_m$ the current is assumed to be drift current only and proportional to n(xm)* E ($x_m$), where one assumes that $x_m \approx 0$ or $E(0) \approx E(x_m)$. Due to this last assumption, the influence of space charge on the value of applied voltage is taken into account only beyond the contact region ($x > x_m$) and the high charge density at the metal/semiconductor interface is neglected (as the contact region is lumped into a single point). For single carrier devices the semi-analytical model shown above (eq 3-5) is similar to the "standard" numerical semiconductor device models [7-9].

## *III.C.  Explicit model*

### III.C.1.        Semiconductor device model

In this paper we present results obtained by a self-consistent solution of an explicit model and compare its results to widely used models for charge injection.

The equations describing the model are:

$$-D \cdot \partial n / \partial x - \mu n \cdot \partial \phi / \partial x = J$$                      (7)

$$\phi = \phi_{SC} + \phi_{image} + \phi_{Applied} \tag{8}$$

$$\partial^2 \phi_{SC}(x)/\partial x^2 = q/(\varepsilon \varepsilon_0) \tag{9}$$

$$\phi_{image}(x) = -q/(8\pi\varepsilon\varepsilon_0 x) + \phi_{barrier} \tag{10}$$

where n is the charge density, D is the diffusivity constant, μ is the mobility, $\phi_{SC}$ is the potential caused by space distribution of charge carriers , $\phi_{image}$ is the image force potential at the contact and $\phi$ is the total potential experienced by the carriers. We compare this model to three other models. To make the comparison simpler we do not account for the field dependence of the mobility[9].

Our numerical simulation is based on solving equation 7 using the exponentially fitted finite difference solution method as outlined in [20,21] and in the appendix. To illustrate the actual meaning of lumping the contact region into single point in space (eq. 4 and 6) we plot in Figure 2 the electronic potential as calculated by "standard" (lumped-contact) models [7-9] along with the explicit model presented here. In these calculations the device length is assumed to be 100nm, the total DOS $N_0 = 10^{20} cm^{-3}$, the mobility $\mu = 10^{-6} cm^2 v^{-1} sec^{-1}$, and the contact barrier is 0.2eV and 0.3eV for Figure 2A and Figure 2B, respectively. The dashed line was derived using lumped-contact model[7-9] and the full line using the explicit model described here. We note that in the lumped-contact models a sizeable region, between x=0 and x=$x_m$, is pushed out of the device (pushed to the left in Figure 2).

In order to examine the change in the physical picture induced by neglecting the contact region we plot in Figure 3 the charge distribution calculated for the same contact barriers as in Figure 2 and net applied voltage ($V_{Appl}$-$V_{bi}$) of 0.5V. This figure illustrates few points. First, for injection-barrier of 0.2eV the charge density at the bulk is almost identical between models and hence one would expect a similar I-V relation (bulk limited). Second, for injection-barrier above 0.2eV (see 0.3eV) the lumped-contact models tend to over estimate the voltage-induced barrier-lowering resulting in a significantly higher charge density in the bulk (hence higher currents). Third, even for cases where injection-contact barrier plays a role the image force still induce a high charge density close to the contact (metal/polymer) interface, as calculated by the explicit model. The neglect of this high-density region of 5-10 nm and its space charge is responsible for the over-estimate of the voltage induced barrier-lowering[22]. We also note that in order to account for high charge density at the interface there is no need to invoke extrinsic traps or defects.

The above effects also manifest themselves in the I-V characteristics of the device. Figure 4A compares simulation results of current–voltage device characteristics to those of the (semi) analytical predictions (lumped-contact models). As expected for low injection-barrier cases SCLC model is a reasonable approximation and for the 0.2eV barrier all three models effectively coincide. For higher injection-barrier, as 0.4eV, the role of the contact region has to be explicitly taken into account, especially at low applied voltages. At high voltages the main draw-back of the lumped-models ("standard") is that they tend to over estimate the voltage-induced barrier lowering effect resulting in an over-estimate of the charge density in the bulk and hence in the current density. This is shown clearly in Figure 4B which shows on linear scale the I-V curves for the 0.6eV barrier case.

## III.C.2.        Accounting for the Gaussian DOS

The advantage of the explicit model described here is that once the contact is made part of the transport equations it becomes possible to account for unique properties associated with organic materials. The most common property is that of disorder and the Gaussian distribution of the density of states (DOS)[4]. It has recently been shown that within the semiconductor transport equation framework the Gaussian DOS results in the mobility ($\mu$) and diffusivity (D) not being related through the classical Einstein relation (D/$\mu$=kT/q) but rather through a generalized relation of the form D/$\mu$=$\eta$•kT/q [23]. Where $\eta$ is a function of both the charge density (n) and the width of the Gaussian DOS ($\sigma$) (see Figure 5). More details of the derivation of the generalized Einstein relation can be found in reference [23]. Note that $\eta$ is strongly dependent on the disorder parameter especially at high charge densities, i.e. this phenomenon should affect the transport at the contact region ($0<x<x_m$) where the density is high (see Figure 3). Mathematically speaking, one should find the self-consistent solution of the following, slightly modified, continuity equation:

$$-\eta_{(n,\sigma)} \cdot \frac{kT}{q} \cdot \mu \cdot \partial n / \partial x - \mu n \cdot \partial \phi / \partial x = J \tag{11}$$

The accounting for $\eta_{(n)}$ requires a modification of the numerical method. For a fine enough grid, such that the density (n) does not change by more then an order of magnitude between cells, one can still make use of the exponentially fitted finite difference scheme [20,21]. In this case the accounting for $\eta_{(n)}$ in the numerical code is made trivial (see appendix). Figure 6 shows the calculated charge density distribution for several disorder parameters ($\sigma$). In the calculation the energy difference between the Gaussian center and the metal work function is fixed at 0.5eV and the net applied voltage (V-$V_{bi}$) is 2V. As the Gaussian width becomes larger there are more transport-states available close to the energy of the metal work-function and hence the injection barrier becomes effectively smaller [15,24]. Figure 7 shows the calculated current voltage relations for the cases shown in Figure 6. The dependence of the mobility ($\mu$) on the disorder ($\sigma$) is not included and hence only the functional form of the curves is important. One should keep in mind that disorder also reduces the mobility and hence the curves in Figure 7 would be shifted slightly downwards ($\mu \propto \exp\left(-\frac{2}{3}\sigma)^2\right)$)[4]. As expected, Figure 7 shows that as $\sigma$ goes up, for a fixed energy difference between the Gaussian center and the metal, the I-V curve tends towards the SCLC functional form. This is consistent with the reduction of the effective barrier as discussed in the context of Figure 6.

## IV.   Conclusions

In conclusion, we have presented self consistent analysis of charge injection and transport in low mobility disordered materials. It was found that incorporating the contact region into the transport model is important to properly account for the contact phenomena. This model shows that a high charge density near the metallic interface is due to the image-force potential and does not require the addition of extrinsic trap states or defects. Moreover, it makes it possible to account for unique features associated with organic materials, as disorder and Gaussian DOS which are known to affect the injection process, all within a conventional semiconductor device model framework. We emphasize that all these effects are entered into the model through a single parameter ,$\eta$ [23], and hence can be added to any semiconductor device model simulator (see appendix).

We expect that the method described here will make it possible to better simulate, design, and manufacture state of the art LEDs that can operate at low voltages and potentially have a fast switching time.

**Acknowledgements**

We acknowledge fruitful discussions with Y. Roichman. We acknowledge support by the EU through contract no. G5RD-CT-2001-00577 OPAMD

## Appendix A – discretization of the continuity equation

The simulation program solves the continuity equation:

$$D\frac{\partial n}{\partial x} + \mu n \frac{\partial \phi}{\partial x} = J$$

where $\Phi(x)$ is a joint potential of space charge induced field, the image potential near the contact and the applied voltage. The charge carriers are assumed to be thermalized at the first hop so that one can assume the concentration in the vicinity of the metallurgic junction to be in equilibrium with metal electrons. If $\Phi(x)$ and $J$ are known one can derive an analytical solution of continuity equation for carrier concentration:

$$n(x) = N_o \exp(-\frac{\mu}{D}\phi(x)) - \frac{J}{D}\exp(-\frac{\mu}{D}\phi(x))\int_0^x \exp(\frac{\mu}{D}\phi(x'))dx'$$

The above representation clearly shows the importance of the ration $\mu/D$. For the numerical solution we apply a discretisation scheme:

$$n_i = n_{i-1}\exp(-\frac{\mu}{D}[\phi_i - \phi_{i-1}]) - \frac{J}{D}\exp(-\frac{\mu}{D}[\phi_i - \phi_{i-1}])\int_{x_{i-1}}^{x_i}\exp(\frac{\mu}{D}\phi(x'))dx'$$

Writing an analogous expression for the next mesh interval and expressing $J$ through $n(x)$ one can arrive at a scheme that contains the carrier concentration only:

$$J = D\left\{\frac{n_{i-1} - n_i\exp(\frac{\mu}{D}[\phi_i - \phi_{i-1}])}{\int_{x_{i-1}}^{x_i}\exp(\frac{\mu}{D}\phi(x'))dx'}\right\} = D\left\{\frac{n_i - n_{i+1}\exp(\frac{\mu}{D}[\phi_{i+1} - \phi_i])}{\int_{x_i}^{x_{i+1}}\exp(\frac{\mu}{D}\phi(x'))dx'}\right\}$$

by rearrangement of terms when assuming $\dfrac{\mu}{D} = \dfrac{q}{kT}$ we arrive at the following discretization scheme[20,21]:

$$\frac{n_{i-1}}{\int_{x_{i-1}}^{x_i}\exp(\frac{q}{kT}\phi(x'))dx'} - n_i\left\{\frac{\exp(\frac{q}{kT}[\phi_i - \phi_{i-1}])}{\int_{x_{i-1}}^{x_i}\exp(\frac{q}{kT}\phi(x'))dx'} + \frac{n_i}{\int_{x_i}^{x_{i+1}}\exp(\frac{q}{kT}\phi(x'))dx'}\right\} + \frac{n_{i+1}\exp(\frac{q}{kT}[\phi_{i+1} - \phi_i])}{\int_{x_i}^{x_{i+1}}\exp(\frac{q}{kT}\phi(x'))dx'} = 0$$

In the generalized Einstein relation case $\frac{\mu}{D} = \frac{q}{\eta kT}$ and hence the discretization scheme is

written as:

$$\frac{n_{i-1}}{\int\limits_{x_{i-1}}^{x_i} \exp(\frac{q}{\eta_{i-1/2}kT}\phi(x'))dx'}\eta_{i-1/2} - n_i \left\{ \frac{\exp(\frac{q}{\eta_{i-1/2}kT}[\phi_i - \phi_{i-1}])}{\int\limits_{x_{i-1}}^{x_i} \exp(\frac{q}{\eta_{i-1/2}kT}\phi(x'))dx'}\eta_{i-1/2} + ... \right.$$

$$\left. ... \frac{n_i}{\int\limits_{x_i}^{x_{i+1}} \exp(\frac{q}{\eta_{i+1/2}kT}\phi(x'))dx'}\eta_{i+1/2} \right\} + \frac{n_{i+1}\exp(\frac{q}{\eta_{i+1/2}kT}[\phi_{i+1} - \phi_i])}{\int\limits_{x_i}^{x_{i+1}} \exp(\frac{q}{\eta_{i+1/2}kT}\phi(x'))dx'} = 0$$

The above scheme is valid only for fine enough grid so that $\eta$ can be assumed to be constant between the mesh points (namely the charge density difference between adjacent points is below a factor of 3).

For any given electronic potential distribution the equation above can be solved to yield

the charge distribution. To account for the self-induced potential (space charge effects)

we solve also the poison equation in a self consistent manner. The algorithm is based on iterative solution until solution convergence is achieved.

## REFERENCES

[1]     N. F. Mott and R. W. Gurney, *Electronic processes in ionic crystals* (Oxford university press, London, 1940).

[2]     M. Pope and C. E. Swenberg, *Electronic Processes in Organic Crystals* (Clarendon Press, Oxford, 1982).

[3]     R. H. Friend, R. W. Gymer, A. B. Holmes, J. H. Burroughes, R. N. Marks, C. Taliani, D. D. C. Bradley, D. A. Dossantos, J. L. Bredas, M. Logdlund, and W. R. Salaneck, Nature **397,** 121-128 (1999).

[4]     M. Van der Auweraer, F. C. Deschryver, P. M. Borsenberger, and H. Bassler, Advanced Materials **6,** 199-213 (1994).

[5]     H. Scher, M. F. Shlesinger, and J. T. Bendler, Physics Today **44,** 26-34 (1991).

[6]     E. M. Conwell and M. W. Wu, Applied Physics Letters **70,** 1867-1869 (1997).

[7]     G. G. Malliaras and J. C. Scott, J. Appl. Phys. **85,** 7426-7432 (1999).

[8]     P. S. Davids, I. H. Campbell, and D. L. Smith, J. Appl. Phys. **82,** 6319-6325 (1997).

[9]     D. J. Pinner, R. H. Friend, and N. Tessler, J. Appl. Phys. **86,** 5116-5130 (1999).

[10]    A. J. Campbell, D. D. C. Bradley, H. Antoniadis, M. Inbasekaran, W. S. W. Wu, and E. P. Woo, Appl. Phys. Lett. **76,** 1734-1736 (2000).

[11]    A. Kraft, G. A.C., and H. A.B., Angew. Chem. Int. Ed. **37,** 402-428 (1998).

[12]    M. A. Baldo and S. R. Forrest, Physical Review B **6408,** 5201-+ (2001).

[13]    J. M. Lupton, V. R. Nikitenko, I. D. W. Samuel, and H. Bassler, Journal of Applied Physics **89,** 311-317 (2001).

14      J. C. Scott and G. G. Malliaras, Chemical Physics Letters **299,** 115-119 (1999).

15      V. I. Arkhipov, U. Wolf, and H. Bassler, Phys. Rev. B **59,** 7514-7520 (1999).

16      E. Tutis, M. N. Bussac, B. Masenelli, M. Carrard, and L. Zuppiroli, Journal of
         Applied Physics **89,** 430-439 (2001).

17      D. F. Blossey, Phys. Rev. B **9,** 5183 (1974).

18      J. Mort, F. W. Schmidlin, and A. I. Lakatos, Appl. Phys. Lett. **42,** 5761-5763
         (1971).

19      B. Masenelli, D. Berner, M. N. Bussac, F. Nuesch, and L. Zuppiroli, Applied
         Physics Letters **79,** 4438-4440 (2001).

20      S. Selberherr, *Analysis and Simulation of Semiconductor Devices* (Springer-
         Verlag, Vien, 1984).

21      M. S. Mock, *Analysis of Mathematical Models of Semiconductor Devices* (Boole
         Press, Dublin, 1983).

22      Y. Preezant, Y. Roichman, and N. Tessler, J. Phys. Cond. Matt. (to be published).

23      Y. Roichman and N. Tessler, Applied Physics Letters **80,** 1948-1950 (2002).

24      V. I. Arkhipov, E. V. Emelianova, Y. H. Tak, and H. Bassler, Journal of Applied
         Physics **84,** 848-856 (1998).

**Figure Caption**

Figure 1:Schematic representation of contact region . $X_m$-coordinate of peak of the image

force potential in the presence of applied field ,$J_t$ – current of carriers thermalized at

the contact region ,$J_h$-current of "hot" carriers that success to overcome the peak

ballisticaly

Figure 2 : Typical picture of band bending. Barrier height is 0.2eV (A) and 0.3eV (B).

The applied voltage is 0.5V beyond the flat band condition (i.e. V ~ 2.5V). The dashed line was calculated using the "standard" (lumped) model and the full line using our explicit model.

Figure 3 : Charge density distribution for a contact injection-barrier of 0.2eV (A) and

0.3eV (B). The dashed line was calculated using the "standard" (lumped) model and the full line using our explicit model.

Figure 4 (A) Current density as function of mean field for 100nm long device and varying

injection-barrier of 0.2eV, 0.4eV, and 0.6eV and device length of 0.1 micron . (B)

Current density for 0.6eV barrier on a linear scale. Standard = Lumped Model,

Explicit = Our Model, SCLC = Space Charge Limited Current as in eq. 3.

Figure 5 : Generalized Einstein relation ($\eta$) as function of charge concentration for

variety of disorder parameter. (calculated based on [23] and $N_0$ is the total DOS)

Figure 6. Charge density distribution

Figure 7. Influence of the disorder on device behavior . Charge distribution and I-V curve

show significant variety for difference disorder in hopping sites energies.

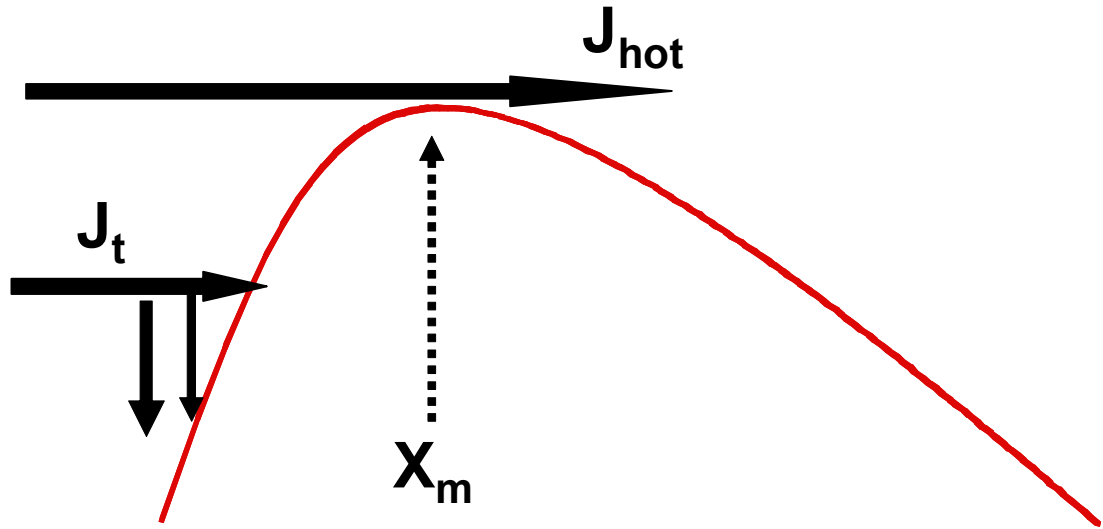

**Figure 1:** Schematic representation of contact region . $X_m$-coordinate of peak of the image force potential in the presence of applied field ,,$J_t$ – current of carriers thermalized at the contact region ,,$J_h$-current of "hot" carriers that success to overcome the peak ballisticaly

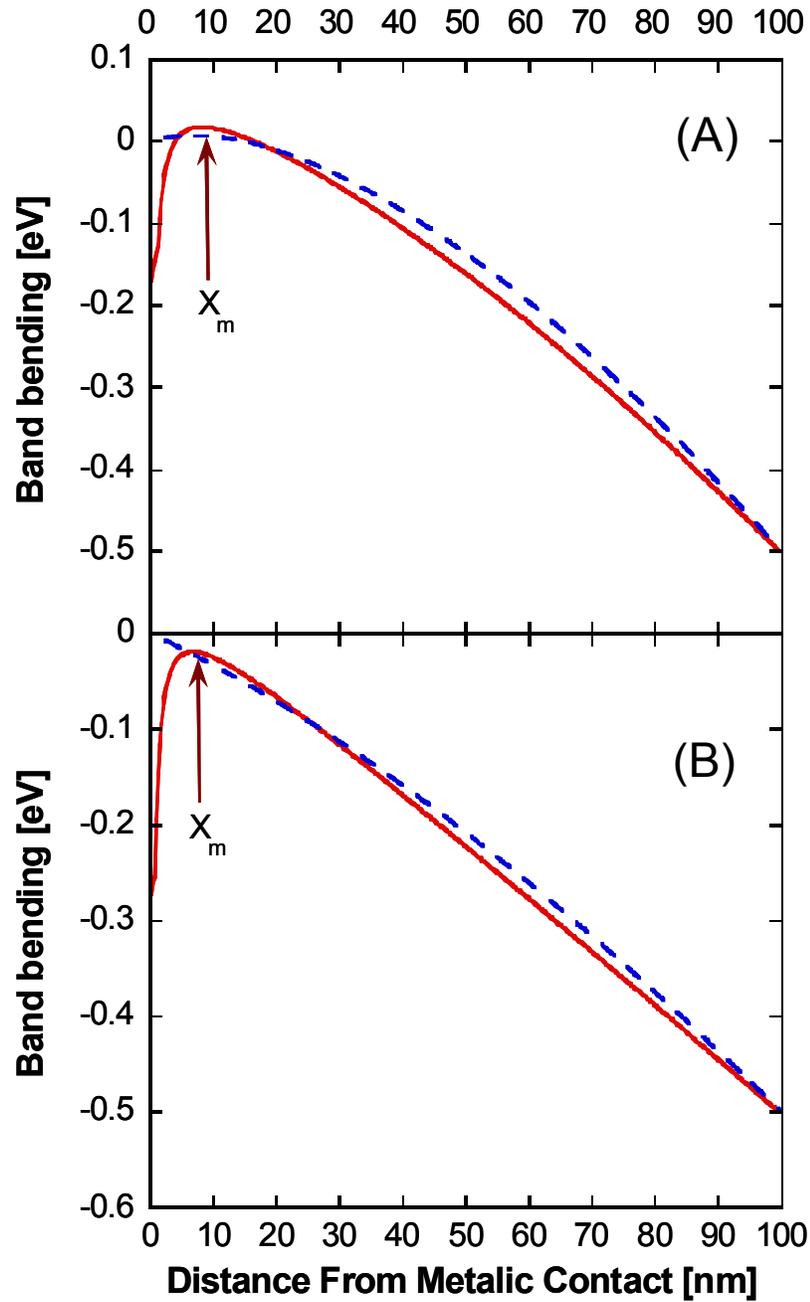

**Figure 2 : Typical picture of band bending. Barrier height is 0.2eV (A) and 0.3eV (B). The applied voltage is 0.5V beyond the flat band condition (i.e. V ~ 2.5V). The dashed line was calculated using the "standard" (lumped) model and the full line using our explicit model.**

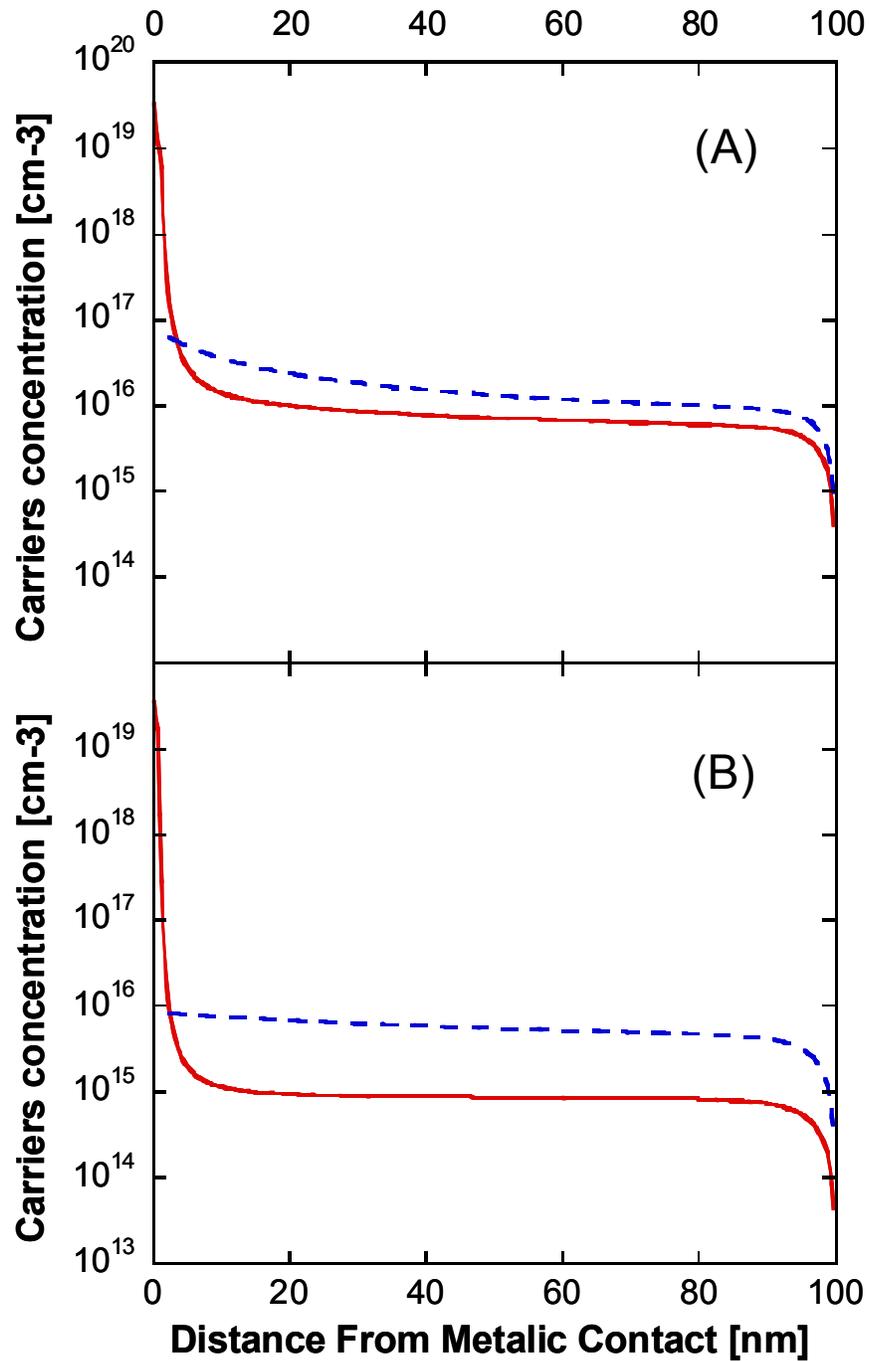

**Figure 3 :** Charge density distribution for a contact injection-barrier of 0.2eV (A) and 0.3eV (B). The dashed line was calculated using the "standard" (lumped) model and the full line using our explicit model.

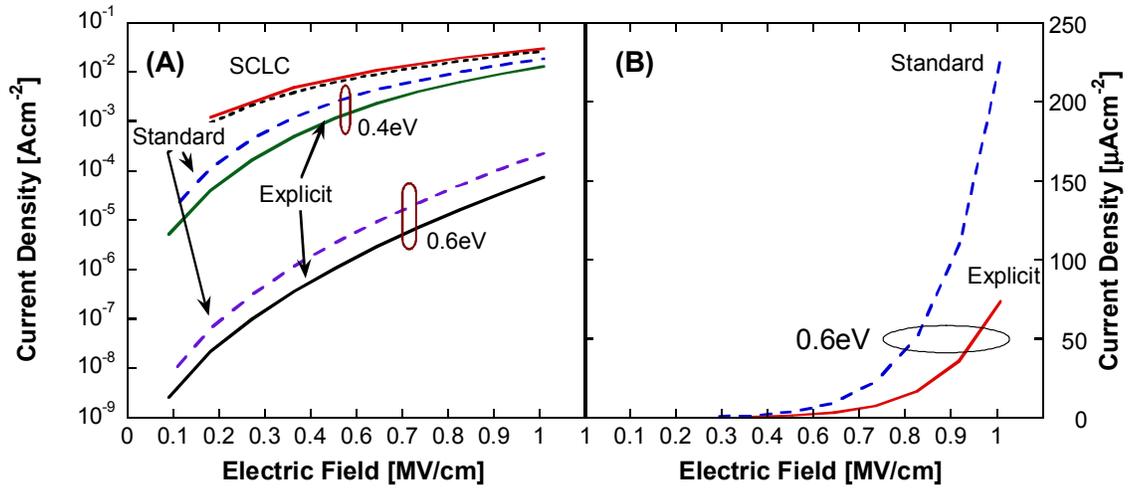

**Figure 4 (A)** Current density as function of mean field for 100nm long device and varying injection-barrier of 0.2eV, 0.4eV, and 0.6eV and device length of 0.1 micron . **(B)** Current density for 0.6eV barrier on a linear scale. Standard = Lumped Model, Explicit = Our Model, SCLC = Space Charge Limited Current as in eq. 3.

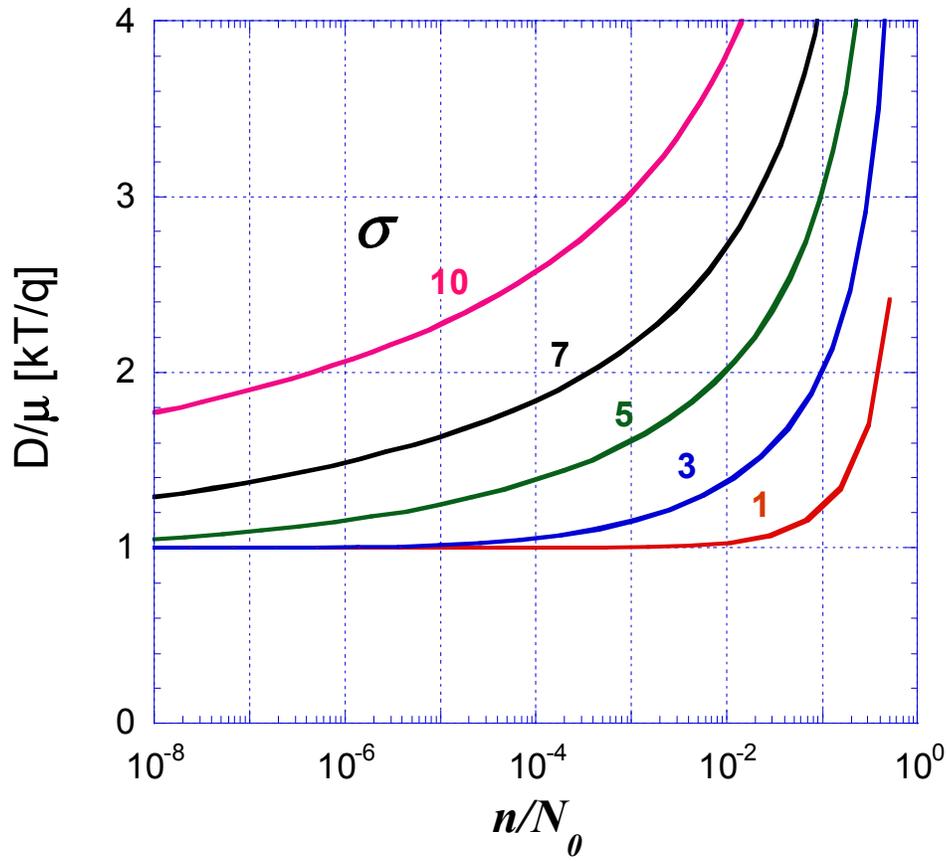

**Figure 5 : Generalized Einstein relation (η) as function of charge concentration for variety of disorder parameter. (calculated based on [23] and N₀ is the total DOS)**

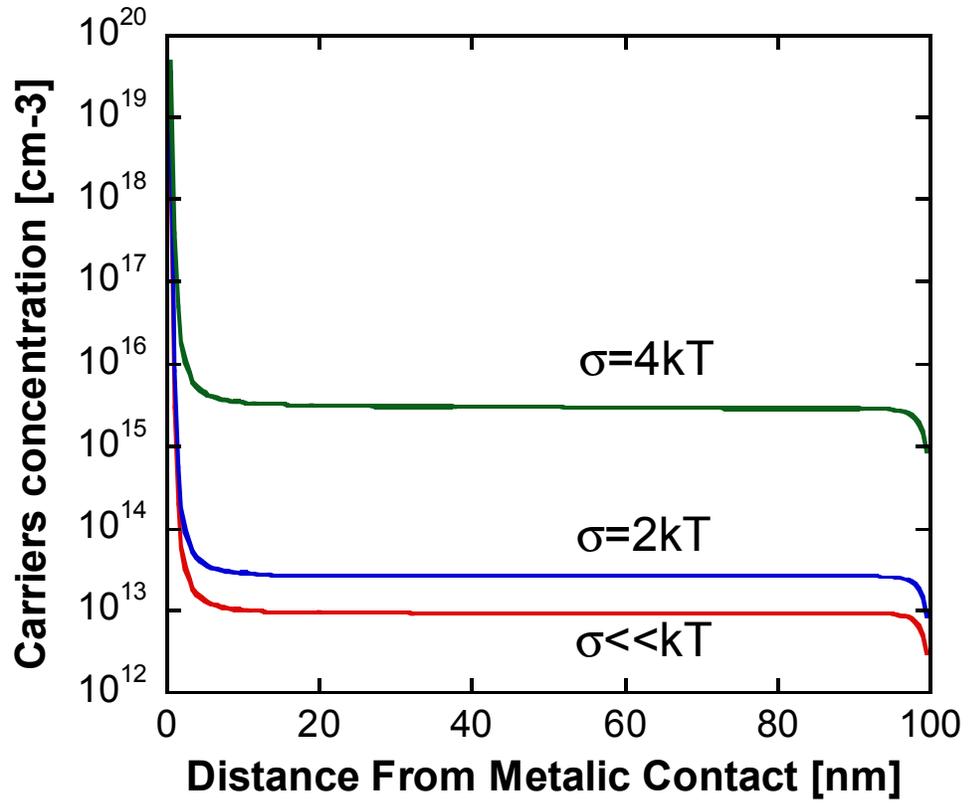

**Figure 6. Charge density distribution**

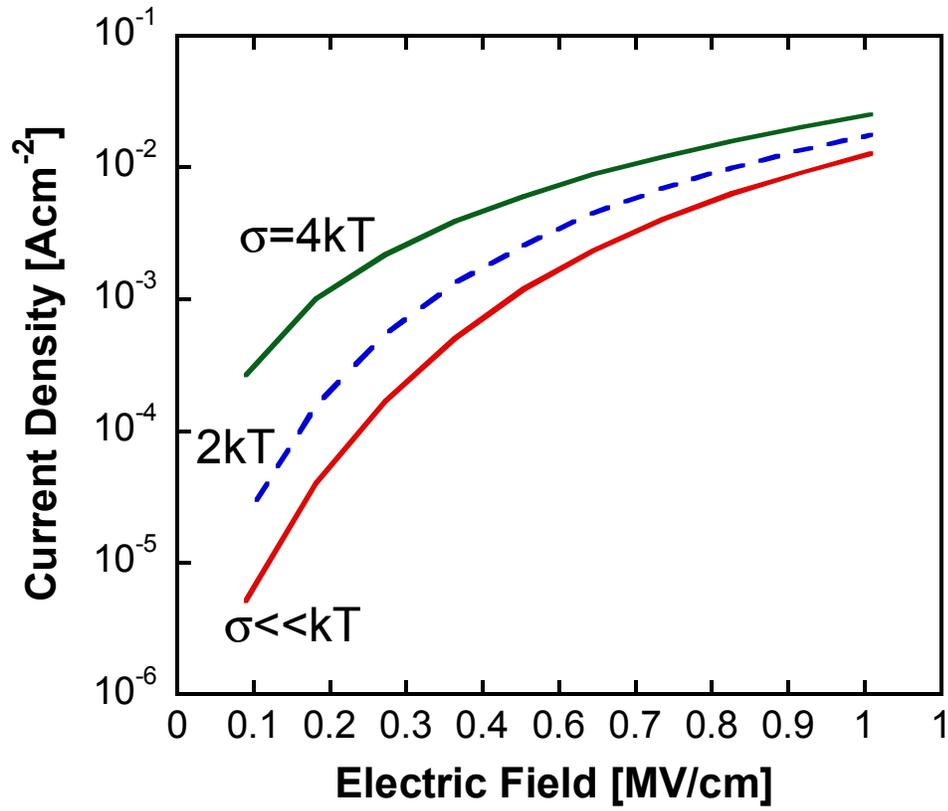

**Figure 7. Influence of the disorder on device behavior . Charge distribution and I-V curve show significant variety for difference disorder in hopping sites energies.**